\documentclass[prb,twocolumn,floatfix,showpacs,superscriptaddress]{revtex4-1} %
\usepackage{amsmath,amsfonts,amssymb,amsthm,mathtools,tabu}
\usepackage{mathptmx}

\newcommand{\ket}[1]{\ensuremath{\left|#1\right\rangle}}

\newcommand{\abs}[1]{\ensuremath{\left| #1 \right|}}

\newcommand{\vk}{\mathbf{k}}
\newcommand{\vp}{\mathbf{p}}
\newcommand{\vq}{\mathbf{q}}
\newcommand{\cG}{{\mathcal  G}}
\newcommand{\cD}{{\mathcal  D}}

\begin{document}

\title{Instability of a dilute Fermi liquid in the presence of forward scattering  }

\author{Yaron Kedem}
\affiliation{Department of Physics, Stockholm University, AlbaNova University Center, 106 91 Stockholm, Sweden}

\begin{abstract}
The stability of a Fermi liquid is analyzed by summing series of diagrams with an interaction mediated by a system close to quantum criticality. The critical temperature and the gap are derived in terms of an effective coupling constant and do not depend on the density of states at the Fermi surface. The forward scattering process is identified as the main pairing mechanism for the case of low, or vanishing, levels of doping. 
\end{abstract}

\maketitle

Cooper instability \cite{Cooper1} is the precursor of the BCS theory of superconductivity \cite{bcs}. While it falls short of providing a full description of the new state of matter, the derivation yields a physical picture, i.e. the Cooper pairs, which is considered valid even after one complement it by introducing a mean field, or a wave function ansatz. Moreover, the critical temperature and the size of the gap can be estimated using fewer assumptions. In particular, they do not depend on the properties of the order parameter, such as symmetries, or in fact even on the existence of order at all. It is thus a good sanity check for any suggested pairing mechanism to see whether the state of a Fermi liquid is indeed unstable when the relevant interaction is present. The nature of the instability might not indicate what the new phase that emerges is, but any region in the phase diagram, where the instability does not exist in, is likely to be a Fermi liquid.  This phenomenon was studied in a wide range of systems, such as composite fermions \cite{coopCompFer}, the Kohn-Luttinger mechanism \cite{coopKohnLutt} and spin liquids \cite{coopSpinLiq}.

The usual derivation of the Cooper instability relies on the adiabatic assumption \cite{migdal}: The energy scale of the interaction is assumed to be much smaller than the Fermi energy so only states close to the Fermi surface are considered. Then, a simplified form of the interaction can be used, with a cutoff on the energy of the electrons, measured from the Fermi energy, and most of the bulk of the Fermi Sea is ignored. However, superconductivity occurs in systems where the energy scale of the modes mediating the interaction is larger than the Fermi energy \cite{sto1,sto1a,dilute1,dilute}. This makes it harder to justify disregarding any electronic states.  Understanding superconductivity in such cases is a long standing \cite{semi,takada,grimaldi} and ongoing \cite{gor2016phonon,ruhman,Gorkov2017,phonon,dirac} effort. It can be valuable to revisit the derivation of the instability, removing  the adiabatic assumption, or at least replacing it with other, more suitable, assumptions.

Originally, acoustic phonons were considered to mediate the attractive interaction. Since then, many other pairing mechanisms were suggested. It is widely believed that modes that are related to a quantum phase transition \cite{sachdevBook} can play an important role in the mechanism, especially in scenarios where the BCS theory is difficult to apply. Recently, the process of forward scattering \cite{forward1,forward2,aniso1,smalq,peak}, in which only small momenta are transferred in the interaction, is thought to contribute significantly to superconductivity in some cases. This process is closely related to the concept of quantum criticality since a system close to a phase transition has a diverging correlation length and thus the interaction mediated by that system can be extremely long ranged. Quantum criticality is also characterized by a vanishing energy scale. This implies the system reacts slowly, a property that is needed for the retardation effect, which allows the effective interaction between electrons to overcome the Coulomb repulsion, in some frequency range, and to yield an attractive coupling.            

Yang and Sondhi \cite{yang} have solved the Cooper problem for long- but finite- ranged interaction by using gradient expansion in momentum space. They showed that indeed a Fermi liquid is unstable and that almost all physical physical quantities are independent of the cutoff. Here we focus on the limit of infinite range, which means some of the expressions below can be obtained by taking that limit $L\rightarrow\infty$ in ref [\onlinecite{yang}]. We show that this limit yields a reasonable physical picture, different from the typical Copper pairing which occurs only on the Fermi surface. Since the instability is derived without the adiabatic assumption, this picture can describe superconductivity at vanishing doping levels. By looking on the propagator of the modes mediating the interaction, the limit is obtained as a consequence of their dispersion relations. 

We start by looking on a simple model. Consider a single electronic band $H_0 = \sum_\vk \xi_\vk c^\dagger_\vk c_\vk$, where $ c^{(\dagger)}_\vk$ is a (creation) annihilation operator of a particle with momentum $\vk$ and $\xi_\vk$ is the energy of that particle. In the ground state $\ket{0}$, all states with $\xi_\vk<0$ are occupied. More explicitly we can write $\xi_\vk= k^2 / 2 m - \mu$, where $m$ is an effective mass and $\mu$ is the chemical potential so $k_F = \sqrt{2 m \mu} $ is the largest momentum state that is occupied. Now consider an interaction,
\begin{align} \label{v1}
H_I = - g \sum_{\vk<\vp}  c^\dagger_\vk c^\dagger_\vp c_\vp c_\vk,
\end{align}
where $g>0$. A state $\ket{\vk_1,\vk_2} =  c^\dagger_{\vk_1} c^\dagger_{\vk_2} \ket{0}$ has an energy $ \xi_{\vk_1} + \xi_{\vk_2} - g$, which can be negative even if $ \xi_{\vk_1}$ and $\xi_{\vk_2}$ are positive. A naive pairing picture would be to claim it is a bound state and that any single particle state, up to an energy $g$ can be paired. Thus, the the Fermi liquid phase is unstable. However, this model has a more severe instability in dimension greater than two. Since $\left[ H_I,H_0 \right] =0$, we can find the eigenstates of the system: they are given by specifying a set $\{ \vk_i \}$ of single particle momenta that are occupied. The contribution of the interaction to the energy is $\propto – g N^2$ where $N$ is the number of occupied states \footnote{Since $H_I$ has a double momentum integration, $\sqrt{g}$ scales inversely with the volume of the system, so one can replace $N$ with the particle density}. At the ground state, small momentum states are occupied and the scaling of the energy contribution from $H_0$ depends on the density of state. In three dimensions, it is $\propto N^{5/3}$ and ground state has a diverging particle density. In two dimensions, the scaling of both contributions is the same and the stability depends on other details.

Indeed, the interaction in Eq. (\ref{v1}) is nonphysical for a number of reasons. For example the limit of infinite range has to be restricted at some scale. Ultimately, that scale would be the system size, but smaller scales can also put bounds on the range of the interaction. More importantly, a bare attractive interaction typically does not yield a thermodynamically stable system (nor does a bare repulsive one). The simple model above disregard the Coulomb repulsion and the positive background charge, as well as any additional screening mechanisms, the dynamics of the modes that mediate the interaction, etc. These effects can be, and usually are, ignored, when the instability is derived. Instead, in the standard treatment, the interaction is taken to affect only states in a given energy range, hence a cutoff is employed. The underlying picture is that the attraction is strongly retarded so in a certain frequency span it can overcome the Coulomb repulsion. Taking into account only these frequencies, by employing the cutoff, a purely attractive interaction is often used, leading to states with negative energies, which are reffered to as bounded, and to an apparent instability. Technically, a perturbative calculation is done and the divergence of the series is seen as evidence for an instability. Here, we adopt some parts of this approach. The attractive interaction is assumed to dominate up to a frequency scale that can be larger than other scales in the problem, such as the Fermi energy or the temperature. The repulsive interaction is neglected but we do not set a cutoff. The perturbative series is calculated, taking into account the frequency dependence of the interaction, and we obtain the parameter space where it is divergent.

The source for the effective interaction between two electrons is a coupling of a single electron to another system, which then mediates the interaction to another electron. The dynamics of the mediating system is a crucial factor for the properties of the resulting effective interaction. The model we consider here is a system close to a quantum phase transition and, in general, our results apply to any such system, provided that it has significant coupling to itinerant electrons at small momentum transfer. An important effect of the proximity to a quantum critical point, in the context of this work, is a diverging correlation length, which implies the Green’s function of the system is highly peaked at zero momentum. In the Matsubara formalism, the Greens function is given by
\begin{align} \label{prop}
\mathcal{D}(\vq,i \omega) = {\Omega \over \omega_\vq^2 + \omega^2 },
\end{align}
where $\Omega$ is some energy scale of the system \footnote{In some models $\Omega=\omega_\vq$ and one can replace  $\Omega \rightarrow \omega_\vq$ throughout the paper.} and $\omega_\vq$ is the frequency of a mode with momentum $\vq$. The important properties of the system, which are a consequence of the proximity to a quantum critical point, are manifested in  $\omega_\vq$, which is not necessarily isotropic, but we assume it has a minimum at $\abs{\vq} =0$. The quantum critical point is characterized by a vanishing gap, i.e. the minimal energy $\omega_{\abs{\vq} =0}$ vanishes. The long range correlations, which typically exists at criticality, implies the curvature $\left.{\partial^2 \omega_\vq \over \partial q_i \partial q_j}\right|_{\abs{\vq} =0} $ is large.  

To be explicit, one can consider a quantum paraelectric system, such as strontium titanate \cite{Itoh1999,rowley}. It was suggested that the ferroelectric modes are the source of superconductivity \cite{jonaPrl} and the connections between the critical behavior of the system and the superconducting transition were  discussed \cite{prbIso}. These ideas have been supported by experimental evidence \cite{stucky,tomioka} and long-range interactions were observed \cite{long}. In ref [\onlinecite{novel}], the model was used to explain superconductivity at vanishing level of doping. There, the propagator in Eq. (\ref{prop}) and the coupling to itinerant electrons $g_\vq$ were derived using the quantum Ising Hamiltonian, with electric dipoles as pseudospins. Here, we do not specify any Hamiltonian for the mediating system, or even consider its degrees of freedom. The only input for the perturbative calculation, are the energies of the electrons, $\xi_\vk$, and the mediating modes $ \omega_\vq $, together with the coupling between them $ g_\vq $. It is assumed $ g_{\vq=0} \ne 0 $, but otherwise $ g_\vq $ can be of a general form.             

\begin{figure}
\centering
\includegraphics[trim=2.5cm 23cm 11.5cm 1.5cm, clip=true,width=0.49\textwidth]{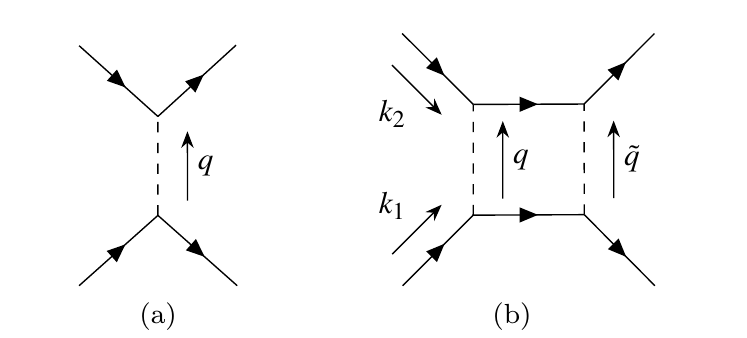}
\caption{(a) The bare scattering process: Fermion propagators, denoted by solid lines, are coupled via modes of the mediating system, with the propagator in Eq. (\ref{prop}), denoted by a dashed line. Each vertex entails a factor of $ g_\vq $. (b) One loop diagram: Two incoming Fermions, with 4-momenta $k_1$ and $k_2$ interact twice, with the 4-momenta $q$ and $\tilde{q}$ being transferred. The outgoing states (not shown) together with the conservation of energy and momentum at each vertex, fixes one variable, say $\tilde{q}$, while the other variable $q$ is integrated over.}
\label{bare}
\end{figure}

The Cooper instability is a divergence of the scattering amplitude of two electrons. We are interested in the scattering due to an effective interaction, mediated by modes with the propagator in Eq. (\ref{prop}). The bare process, shown in Fig \ref{bare} (a), yields 
$V_0 =   \cD\left(q \right)  \abs{g_\vq}^2$,
where we have used the four-component notation $q=(i\omega,\vq)$ for the transferred momentum and energy. The coupling $ g_\vq $ is assumed to be a Fourier transform of a real function implying   $ g_{-\vq} =  g^*_\vq $. Other contributions, which can also be represented as diagrams, will be of higher orders in $\cD\left(q \right)  \abs{g_\vq}^2$. If this factor is small, it might be possible to sum all contributions, or a partial set of them that is deemed significant, to obtain a finite result. Then, it is expected that the ground state would be a Fermi liquid with renormalized parameters. A diverging result is an indication of an instability, implying that our starting point for the perturbative calculation, which is the state with no coupling, is not adiabatically connected to the true ground state when the coupling is present. 

We can start with the second order diagram, shown in Fig \ref{bare} (b), which contains a loop and thus entails an integration over an internal degrees of freedom
\begin{align} \label{two}
 V_1 = \sum_{q}  \abs{g_{\vq}}^2  \abs{g_{\tilde{\vq}}}^2  \cD\left(\tilde{q}\right)   \cD(q) \cG\left(k_1 + q \right) \cG\left(k_2 -q \right), 
\end{align}
where $k_1$ and $k_2$ ($q$ and $\tilde{q}$) are the 4-momenta of the incoming electrons (mediating modes) and $\cG\left(k \right) = 1/(i \omega - \xi_\vk)$ is the electronic Matsubara Green's function. Using the four-component notation means the summation over the internal degrees of freedom involves a sum over Matsubara frequencies and an integral over the Brillouin zone $\sum_q \rightarrow T \sum_m \int d^3q$, where $T$ is the temperature and $m$ is the frequency index $\omega = 2 \pi T m$. The frequency summation can be done exactly and the resulting expressions, shown in the appendix, are rather cumbersome. Here, we focus on two limits, each implying a certain approximation, and obtain much simpler expressions. The limits can be expressed as considering $\omega_\vq$ in Eq. (\ref{prop}) to be very small or very large. In terms of Eq. (\ref{two}), small $\omega_\vq$ implies  $\cD(q)$ is strongly peaked at small $m$, compared to the $m$ dependency of $\cG\left(k_{1,2} \pm q \right)$, and for large $\omega_\vq$ the situation is reversed. Here, large, or small, is mainly with comparison to $T$, which sets the energy scale of the Matsubara frequencies, but in some steps $\xi_\vk$ can also be relevant.
   
We start with the case of large $\omega_\vq$, which is similar to the conventional derivation of superconductivity. Assuming $\cD(q) \simeq \Omega \omega_\vq^{-2}$ is roughly independent of frequency, the sum in Eq. (\ref{two}) is now over the two electronic Greens functions only, which yields    
\begin{align} \label{twoa}
 V_1^I =  \int d^3q  { \abs{g_{\vq}}^2  \abs{g_{\tilde{\vq}}}^2 \Omega^2 \over \omega_\vq^2 \omega_{\tilde{\vq}}^2} {N_F(\xi_{\vk_1 + \vq}) -N_F(-\xi_{\vk_2 - \vq}) \over i (k^0_1+ k^0_2) - (\xi_{\vk_1 + \vq} +\xi_{\vk_2 - \vq})}, 
\end{align}
where $k^0_{1(2)}$ is the Matsubara frequency related to $k_{1(2)}=\left(i k^0_{1(2)},\vk_{1(2)} \right)$ and $N_F(\xi)=1/(e^{\xi/T} +1)$ is the Fermi occupation function. This assumption implies $\omega_\vq \gg T$ but also $\omega_\vq \gg \xi_\vk$, since $ \xi_\vk$ determines the width of $\cG\left(k \right)$ in the frequency domain, and that width is assumed be smaller than the width of $\cD(q)$. This makes it natural to set a cutoff on $\xi_\vk$ that is related to $\omega_\vq$, i.e. the Debye frequency. Our focus is on the forward scattering process and, as we show now, no cutoff is required. Instead, we assume $\omega_\vq^{-2}$ is strongly peaked at $\abs{\vq} =0$. Technically, we replace it with a suitably normalized delta function $\omega_\vq^{-2} \propto \delta(\vq)$, which makes the momentum integral trivial. More specifically, $\omega_\vq^{-2}$ is assumed to be negligible outside a small region around $\abs{\vq} =0$. The other quantities in the integrand are assumed to do not have a strong momentum dependency at that region $\xi_{\vk + \vq} \simeq \xi_{\vk}, \abs{g_{\vq}}^2 \simeq \abs{g_{0}}^2$. The size of that region, which will affect the result, can be incorporated into the normalization of the delta function. We get
\begin{align} \label{v1a}
 V_1^I =   \abs{g_0}^2 { \Omega \over \omega_0^2 } \lambda  {N_F(\xi_{\vk_1 }) - N_F(-\xi_{\vk_2 }) \over i (k^0_1+ k^0_2) - (\xi_{\vk_1} +\xi_{\vk_2})}, 
\end{align}
where 
\begin{align} \label{lambda}
\lambda = \int d^3q \abs{g_{\vq}}^2 { \Omega \over \omega_\vq^2 } \propto \abs{g_0} { \Omega \over \omega_0^2 }
\end{align}
is the coupling constant for the effective interaction. Unlike typical definitions for the coupling constant  \cite{mahan}, which include the electronic density of states at the Fermi surface or the Fermi velocity, according to the definition of Eq. (\ref{lambda}),  $\lambda$ has units of energy. The discrepancy is due to the different nature of the assumptions leading to the definitions. Instead of assuming the interaction occurs only at the Fermi surface and employing a cutoff in energy, we assume the interaction occurs only at small momentum and employ a delta function in momentum. In the scenario we are considering, a low doping level and an interaction mediated by a quantum critical system, the former assumption is unjustified while the latter is. 

We now turn to the case of small $\omega_\vq$. When $\omega_\vq \ll T$, $\cD(q)$ is strongly peaked at $\omega=0$ where $\cD(0,\vq) = \Omega \omega_\vq^{-2}$, while $\cD(\omega \ne 0,\vq) \propto T^{-2}$. To first order in $\omega_\vq / T$ we can neglect all terms in the frequency sum in Eq. (\ref{two}), besides  $\omega=0$, and obtain  
\begin{align} \label{two2}
 V_1^{II} =  \int d^3q \abs{g_{\vq}}^2  \abs{g_{\tilde{\vq}}}^2 { \Omega^2 \over \omega_\vq^2 \omega_{\tilde{\vq}}^2} { T \over (i k^0_1 - \xi_{\vk_1 + \vq})(i k^0_2 - \xi_{\vk_2 - \vq})}. 
\end{align}
The momentum integral is done as before and we get
\begin{align} \label{v1b}
 V_1^{II} =   \abs{g_0}^2 { \Omega \over \omega_0^2 }   { \lambda T \over (i k^0_1 - \xi_{\vk_1})(i k^0_2 - \xi_{\vk_2 })},
\end{align}
where $\lambda$ has the same definition, Eq.  (\ref{lambda}). 

The expressions in Eq. (\ref{v1a}) and (\ref{v1b}) refer to two different limits of  $\omega_0/T$. For any finite value of these parameters, the two contributions are approximations and can be added, $V_1 \simeq V_1^{I} + V_1^{II}$. A large (small) value of $\omega_0/T$ implies $  V_1^{I(II)}$ dominates and also that it is a better approximation. In what follows, we treat the two limits as two separate cases, so each of them can be described with relatively simple expressions. The full expression, including the two contributions and other corrections, is give in the appendix.    

\begin{figure}
\centering
\includegraphics[trim=2.2cm 22.5cm 11.2cm 1.4cm, clip=true,width=0.49\textwidth]{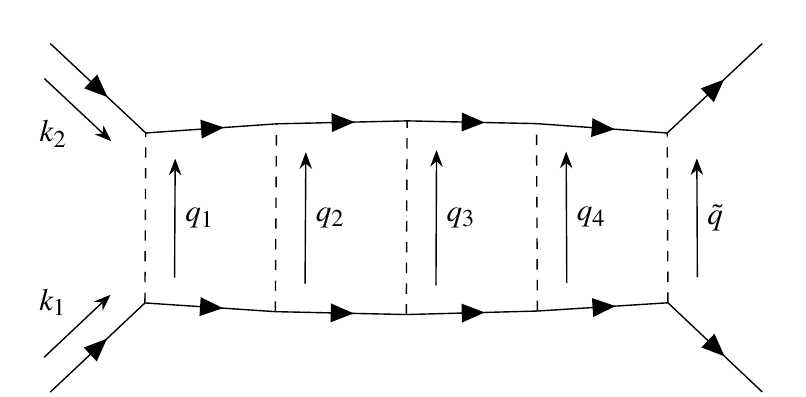}
\caption{A ``Ladder'' diagram with 4 loops, corresponding to $V_4$ given by Eq. (\ref{full}).}
\label{ladder}
\end{figure}

Now we want to study higher order diagrams. We consider the so-called ``ladder diagrams'', as shown in Fig \ref{ladder}, and attempt to sum them up to infinite order. We disregard the crossing diagrams, but do not expect an inclusion of those to change the result qualitatively. The full expression for the effective interaction is $V_\text{tot} =  \sum_{n=0}^\infty V_n$, with
\begin{align} \label{full}
V_n =   \sum_{q_1..q_n} V_0 \prod_{j=1}^n \abs{g_{\vq_j}}^2 \cD(q_j) \cG\left(k_1 + \sum_{i=1}^j q_i\right) \cG\left(k_2 - \sum_{i=1}^j q_i\right). 
\end{align}
where $V_0 =   \cD \left(\tilde{q} \right)  \abs{ g_{\tilde{ \vq }} }^2$ is the bare interaction for the momentum $\tilde{q}$ that is not integrated over. We have already evaluated $V_1$, which is given by Eq. (\ref{v1a}) or Eq. (\ref{v1b}). For a general term $V_n$, the summation over the last variable $q_n$ is similar to the summation we did when evaluating $V_1$, with the mapping  $k_1 \rightarrow k_1 + \sum_{i=1}^{n-1} q_i $ and $k_2 \rightarrow  k_2 - \sum_{i=1}^{n-1} q_i $. In principle, the result would depend on the other 4-momenta, via the new variables $k_1$ and $k_2$. Since we apply a delta function in (spatial) momentum, there is no dependency on $\vq_{j<n}$ and the mapping is the identity $\vk \rightarrow \vk $. Regarding the dependency on the frequency, in Eq. (\ref{v1a}), $V_1$ depends on $i k^0_1 +i k^0_2$, which is invariant under the mapping, i.e. does not depend on $q_{j<n}$. In obtaining Eq. (\ref{v1b}), we have applied a delta function in frequency as well and there is no dependency on $q_{j<n}$ at all\footnote{ $V_0$ can also be considered to depend on $q_{j}$ via the variable $\tilde{q}$, but the same arguments allow us to neglect this dependency when carrying out the summations.}. Thus, we can carry out the summations and get 
\begin{align} \label{vna}
 V^I_n =   \abs{g_0}^2 { \Omega \over \omega_0^2 } \left( \lambda  {N_F(\xi_{\vk_1 }) -N_F(-\xi_{\vk_2 }) \over i (k^0_1+ k^0_2) - (\xi_{\vk_1 } +\xi_{\vk_2})}\right)^n, 
\end{align} 
for $\omega_0/T \gg 1$ and
\begin{align} \label{vnb}
 V^{II}_n =   \abs{g_0}^2 { \Omega \over \omega_0^2 }  \left( { \lambda T \over (i k^0_1 - \xi_{\vk_1})(i k^0_2 - \xi_{\vk_2 })}\right)^n
 \end{align} 
for $\omega_0/T \ll 1$.

In order for the series to converge, the term that is exponentiated to the power of $n$, should have a magnitude smaller than unity. We will now study this term, the one inside the large parenthesis in Eq. (\ref{vna}) and (\ref{vnb}), and its dependency on $T$, $k_{1,2}$ and $\lambda$ to see when the series diverge, i.e. for which parameters the system is unstable. Our interest is not in the usual divergence, when the incoming particles are ``on shell'' $ i k^0 = \xi_\vk$. Instead, we want to study states with a minimal frequency, or vanishing in the case of $T=0$, and see for what momenta the series diverges.

For Eq. (\ref{vna}) the condition is
\begin{align} \label{ca}
\abs{ \lambda { N_F(\xi_{\vk_1 }) - N_F(-\xi_{\vk_2 }) \over \xi_{\vk_1 } +\xi_{\vk_2}}} >1. 
\end{align} 
At $T=0$, the denominator amounts to $(-)1$ when $\xi_{\vk_1 }, \xi_{\vk_2 } (<)> 0$ and vanishes otherwise, i.e. the momenta $\vk_{1,2}$ have to be both inside, or outside, the Fermi volume. It is also requires that the total energy have to be smaller than the coupling energy, $\xi_{\vk_1 } + \xi_{\vk_2 } < \lambda$. Thus, we can expect a gap of size $\lambda$ with states above it still representing single quasi-particles while for lower energies other type of states emerge. At higher temperatures the denominator in Eq. (\ref{ca}) decreases and so does the gap. The critical temperature, is obtained by taking the limit  $\xi_{\vk_1 },\xi_{\vk_2 } \rightarrow 0$. Assuming $\xi_{\vk_1 }=\xi_{\vk_2 } = \xi \ll T$, we have $N_F(\xi) - N_F(-\xi) \simeq \xi / (2 T)$ and Eq. (\ref{ca}) is reduced to $T_c>T $, where
\begin{align}
T_c^I=\lambda/4
\end{align} 
is the critical temprature.

In deriving Eq. (\ref{vnb}), we assumed $\omega_0/T \ll 1$ so one cannot set $T=0$. Instead we set $  k^0_{1,2} = \pi T$ and obtain the condition,
\begin{align} \label{cb}
\abs{ \lambda T \over (i \pi T - \xi_{\vk_1})(i \pi T - \xi_{\vk_2 })} > 1,
 \end{align}
for the series to diverge. For any finite $T$, Eq. (\ref{cb}) can be solved to obtain the values of $\xi_{\vk_1 }$ and $\xi_{\vk_1 }$ for which it is satisfied. The largest value 
\begin{align}
\xi_{\vk_1 }^{max} ={1 \over \pi} \Re{\sqrt{\lambda^2 - \pi^4 T^2}},
\end{align}
which is obtained by setting  $\xi_{\vk_2 }=0$, might be interpreted as a gap. Here as well, the critical temperature, 
\begin{align}
T_c^{II}=\lambda/\pi^2,
\end{align} is obtained by taking the limit  $\xi_{\vk_1 },\xi_{\vk_2 } \rightarrow 0$. 

These results indicate that in a certain temperature range, in a certain energy range, the system is unstable and unlikely to be in a Fermi liquid state. The nature of the divergence, which is due to the scattering process of two particles, is often interpreted as showing a new type of state where the particles are paired. However, the derivation itself does not explicitly imply anything about the new state that might emerge. Only when one employs some mean field, or a wavefunction ansatz, can the properties of the new state, such as symmetry breaking or even superconductivity itself, be discussed. In fact, these properties are typically introduced by assumption, and the main support for their existence is the self consistency of the solutions.

\begin{figure}
\centering
\includegraphics[trim=2.2cm 21.3cm 11.2cm 1.5cm, clip=true,width=0.49\textwidth]{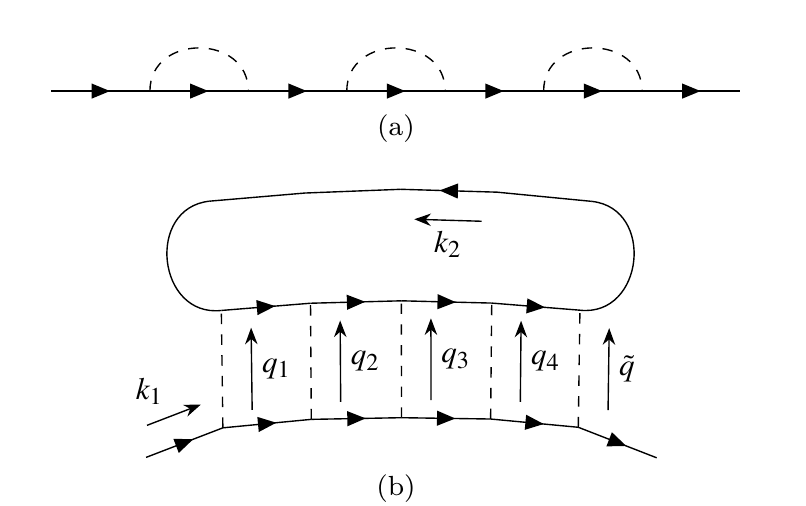}
\caption{Single particle diagrams: (a) The process of one electron interacting with a mode of the mediating system, corresponding to a term with $n=3$ in the sum in Eq. (\ref{full1}). (b) A physical electrons interacting with a virtual one via modes of the mediating system, corresponding to a term with $n=4$ in the sum in Eq. (\ref{G2}).}
\label{self}
\end{figure}

It is possible that the pairing picture arises simply because we focused on the interaction between two particles. Let us now consider other series of diagrams, related to one particle processes, for example the one shown in Fig. \ref{self} (a). As before, we do not include crossing diagrams. The full expression for the series is given by
\begin{align} \label{full1}
G_\text{eff} =  \sum_{n=0}^\infty  \left(  \cG\left(k\right)\sum_{q}  \abs{g_{ \vq} }^2 \cD\left( q \right)  \cG\left(k - q \right) \right)^n \cG\left(k\right).
\end{align}
As before, we consider the forward scattering process, making the integral over the internal variable $\vq$ trivial. The frequency summation is done for the two limits and yields  
\begin{align} \label{gna}
 G^I_n =   {1 \over i k - \xi_{\vk }} \left( \lambda  {N_F(\xi_{\vk }) \over i k - \xi_{\vk }}\right)^n, 
\end{align} 
for $\omega_0/T \gg 1$ and
\begin{align} \label{gnb}
 G^{II}_n =   {1 \over i k - \xi_{\vk }} \left( { \lambda T \over (i k - \xi_{\vk})^2 }\right)^n
 \end{align} 
for $\omega_0/T \ll 1$. 
The divergence of this series indicates the Fermi liquid is unstable in a certain parameter range, similar to the results before. The previous conclusion, namely that a new many body state will emerge in this parameter range, is supported by this calculation as well. However, the physical picture of pairing is irrelevant here. 

Another series of diagrams, shown in Fig. \ref{self} (b), yields  
\begin{align} \label{G2}
G_\text{eff} =  \sum_{n=0}^\infty \sum_{k_2} V_n(k_1,k_2) \cG\left(k_2\right),
\end{align}
where $V_n(k_1,k_2)$ is given by Eq. (\ref{full}). The frequency summation, related to the variable $k_2$, can be done exactly when $v_n$ is given by Eq. (\ref{vna}) or Eq. (\ref{vnb}). The results include derivatives $\partial^{n-1} N_F(\xi)/\partial^{n-1} \xi$, which have significant support only close to the Fermi surface. The momentum integral, which is not restricted to the forward scattering process in this case, can also be done analytically since the only dependency is via $N_F(\xi_\vk)$. Terms that include the number, or density, of electrons $\rho = \int N_F(\xi_\vk) d\vk$, are suppressed in the low doping regime. 

It should be emphasized that the calculations above should not by any means be understood as a exhaustive list of diagrams. Neither do we argue that these represent the most important physical processes for the formation of a new phase, superconducting or otherwise. The aim of this work is to point out some processes that induce an instability, which can be obtained with a high level of analytical control.  

Unlike the usual derivation of the Cooper instability, our results do not depend strongly on the chemical potential $\mu$, or the density of states at the Fermi level $N(0)$. In fact, the approximation we used, $\xi_{\vk +\vq} \simeq \xi_{\vk}$ for small $q$, is better for small $\vk$, which is the case, at the Fermi surface, when $\mu$ and $N(0)$ are small. This implies that forward scattering process can dominate the paring mechanism at low, and even vanishing, levels of doping.

We have showed that a Fermi liquid is unstable in the presence of an attractive forward scattering process. Interactions mediated by a system close to a QCP are inherently long ranged and thus are likely to be dominated by forward scattering. In the quantum critical regime, the zero frequency term, in the summation over Matsubara frequencies, can dominate the sum and yields a qualitatively different form for the effective interaction. The calculation of instability indicates there is an energy gap and a critical temperature, above which the Fermi liquid might be stable. The results do not indicate what is the new many body state that can replace the Fermi liquid.

{\it Acknowledgments.---} This work was supported by the Swedish research council (VR) and the Wallenberg Academy Fellows program of the Knut and Alice Wallenberg Foundation. I am grateful to C. Triola and I. Sandalov for useful discussions. 

\begin{widetext}

\appendix
\section*{Appendices}
\renewcommand{\thesubsection}{\Alph{subsection}}
 
\subsection{Frequency summation}

We want to perform the sum over Matsubara frequencies of a one loop expression, given in Eq. (3) and shown in Fig. 2, in the main text . That is, we want to calculate
\begin{align}
V=&\sum_{n}  \cD(\vq, i \omega_n) \cD\left(  \tilde{ \vq}, i  \tilde{\omega}_n \right) \cG\left(\vk_1 + \vq,i k^0_1+ i \omega_n \right)  \cG\left(\vk_2 - \vq,i k^0_2 - i \omega_n \right) \nonumber \\
 =&\sum_{n} {\Omega \over \omega_\vq^2 + \omega_n^2 }  {\Omega \over \omega_{\tilde{ \vq}}^2 + (\tilde{\omega}_n )^2 }{1 \over i k^0_1+ i \omega_n - \xi_{\vk_1 + \vq}} {1 \over i k^0_2 - i \omega_n - \xi_{\vk_2 - \vq}}
\end{align}
where $ \tilde{\omega}_n =  \omega_n + i \delta k$, the frequency of the second propagator, is given by $ \omega_n$, which is summed over, and $\delta k$ given by the total change in frequency in the final states, which is not summed over. $ \omega_n$ (and also $ \tilde{\omega}_n$) is bosonic so the we can evaluate the sum using the integral  $I=\int dz N_B( z) f(z)= 0$, where $N_B( z)$ is the Bose-Einstein distribution and   
\begin{align}
f(z)  = {\Omega^2 \over (\omega_\vq +z)(\omega_\vq - z)  ( \omega_{\tilde{\vq}} + i \delta k +z) ( \omega_{\tilde{\vq}} - i \delta k -z)(i k^0_1 - \xi_{\vk_1 + \vq} +z) ( i k^0_2 - \xi_{\vk_2 - \vq}-z)}.
\end{align}
The poles, $z$ and residues $R$ of the integral are given in Table \ref{poles}. For the two fermionic poles we have used the Fermi-Dirac distribution $ N_F(  \xi_{\vk} )= N_B( \xi_{\vk}+ ik)$. 
\begin{table}[b]
\caption{\label{poles} poles and residues of $N_B( z) f(z)$}
\begin{tabular}{|l|c|}\hline
$z$&$R$\\  \hline
$ \omega_\vq$ &  ${ N_B( \omega_\vq)  \Omega^2 \over (2 \omega_\vq )  ( \omega_{\tilde{ \vq}} + i \delta k +\omega_\vq) ( \omega_{\tilde{ \vq}} - i \delta k -\omega_\vq) (i k^0_1 - \xi_{\vk_1 + \vq} +\omega_\vq)( i k^0_2 - \xi_{\vk_2 - \vq}-\omega_\vq)}$ \nonumber \\ \hline
 $- \omega_\vq$ &     ${ - N_B( - \omega_\vq)\Omega^2 \over (2 \omega_\vq)  ( \omega_{\tilde{ \vq}} + i \delta k -\omega_\vq) ( \omega_{\tilde{\vq}} - i \delta k +\omega_\vq) (i k^0_1 - \xi_{\vk_1 + \vq} -\omega_\vq)( i k^0_2 - \xi_{\vk_2 - \vq}+\omega_\vq)}$ \nonumber \\ \hline
$- i \delta k +  \omega_{\tilde{\vq}}$ &${ N_B( \omega_{\tilde{\vq}} )  \Omega^2 \over (\omega_\vq - i \delta k +  \omega_{\tilde{\vq}})(\omega_\vq + i \delta k -  \omega_{\tilde{\vq}}) (i k^0_1 - \xi_{\vk_1 + \vq} - i \delta k +  \omega_{\tilde{\vq}}) (2 \omega_{\tilde{\vq}} )  ( i k^0_2 - \xi_{\vk_2 - \vq}+ i \delta k -  \omega_{\tilde{\vq}})}$ \nonumber \\ \hline
 $- i \delta k -  \omega_{\tilde{\vq}}$ &${- N_B(  -  \omega_{\tilde{\vq}} )   \Omega^2 \over (\omega_\vq - i \delta k -  \omega_{\tilde{\vq}})(\omega_\vq  + i \delta k +  \omega_{\tilde{\vq}}) (i k^0_1 - \xi_{\vk_1 + \vq} - i \delta k -  \omega_{\tilde{\vq}}) ( 2 \omega_{\tilde{\vq}}) ( i k^0_2 - \xi_{\vk_2 - \vq}+ i \delta k +  \omega_{\tilde{\vq}})}$ \nonumber \\ \hline
  $\xi_{\vk_1 + \vq} - i k^0_1$  &${- N_F(  \xi_{\vk_1 + \vq} )   \Omega^2 \over (\omega_\vq +\xi_{\vk_1 + \vq} - i k^0_1)(\omega_\vq - \xi_{\vk_1 + \vq} + i k_1)  ( \omega_{\tilde{\vq}} + i \delta k +\xi_{\vk_1 + \vq} - i k^0_1) ( \omega_{\tilde{\vq}} - i \delta k -\xi_{\vk_1 + \vq} + i k^0_1) ( i k^0_2 - \xi_{\vk_2 - \vq}- \xi_{\vk_1 + \vq} + i k^0_1)}$ \nonumber \\ \hline
 $i k^0_2 - \xi_{\vk_2 - \vq}$ & ${ N_F( - \xi_{\vk_2 - \vq} )   \Omega^2 \over (\omega_\vq +i k^0_2 - \xi_{\vk_2 - \vq})(\omega_\vq - i k^0_2 + \xi_{\vk_2 - \vq}) ( \omega_{\tilde{\vq}} + i \delta k +i k^0_2 - \xi_{\vk_2 - \vq}) ( \omega_{\tilde{\vq}} - i \delta k -i k^0_2 + \xi_{\vk_2 - \vq})  (i k^0_1 - \xi_{\vk_1 + \vq} +i k^0_2 - \xi_{\vk_2 - \vq})}$ \nonumber \\ \hline
  \end{tabular}
\end{table}

The are 3 energy scales involved, $\omega_\vq$, $ \xi_{\vk}$ and also $T$, since the fermionic frequencies, $ik^0_{1,2} \propto T$. The transfer of frequency $\delta k$ is bosonic and is assumed to vanish. In the first case $\omega_\vq \gg \xi_{\vk},T$ so we neglect $\xi_{\vk}$ and $ik^0_{1,2}$, where they are added to  $\omega_\vq$. Then, the residues of the 4 bosonic poles are $O(\omega_\vq^{-5})$ and much smaller the residues of the 2 fermionic poles, which are $O(T^{-1}\omega_\vq^{-4})$. The sum of the 2 fermionic poles yields Eq. (4) in the main text.

In the second case $\omega_\vq \ll T$, we neglect $\omega_\vq$ where it is added to $ik^0_{1,2}$. Then, the residues of the four bosonic poles are $O(\omega_\vq^{-3}T^{-2})$ and much larger than the residues of the two fermionic poles, which are $O(T^{-5})$. Furthermore, $ N_B( \omega_\vq) \sim T/ \omega_\vq$, so sum of the 4 bosonic poles yields Eq. (7) in the main text.

\end{widetext}

\end{document}